\documentclass[%
superscriptaddress,
twocolumn,
showpacs,
amsmath,amssymb,
aps,
prb,
floatfix,
showkeys,
reprint,
]{revtex4-1}

\usepackage[dvips]{graphicx}
\usepackage{dcolumn}
\usepackage{bm}
\usepackage{amsmath}
\usepackage{amssymb}
\usepackage{multirow}
\usepackage{booktabs}
\usepackage{tabularx}
\usepackage[english]{babel}
\usepackage[utf8]{inputenc}
\usepackage[version=4]{mhchem}
\usepackage{enumerate}

\usepackage[normalem]{ulem}

\newcommand{\etal}{\textit{et al.\/}}

\newcommand{\eg}{e.\,g.,}

\DeclareUnicodeCharacter{FB01}{\,}

\begin{document}

\title{First-principles prediction of sub-10 nm skyrmions in Pd/Fe bilayers on Rh(111)}

\author{Soumyajyoti Haldar}
\email[Corresponding author: ]{haldar@physik.uni-kiel.de}
\affiliation{Institute of Theoretical Physics and Astrophysics, University of Kiel, Leibnizstrasse 15, 24098 Kiel, Germany}

\author{Stephan von Malottki}
\affiliation{Institute of Theoretical Physics and Astrophysics, University of Kiel, Leibnizstrasse 15, 24098 Kiel, Germany}

\author{Sebastian Meyer}
\affiliation{Institute of Theoretical Physics and Astrophysics, University of Kiel, Leibnizstrasse 15, 24098 Kiel, Germany}

\author{Pavel F. Bessarab}
\affiliation {School of Engineering and Natural Sciences, University of Iceland, 107, Reykjavik, Iceland}
\affiliation {ITMO University, 197101, St. Petersburg, Russia}

\author{Stefan Heinze}
\affiliation{Institute of Theoretical Physics and Astrophysics, University of Kiel, Leibnizstrasse 15, 24098 Kiel, Germany}

\date{\today}

\begin{abstract}
We show that stable skyrmions with diameters of a few nanometers can emerge in atomic Pd/Fe bilayers on the Rh(111) surface.
Based on density functional theory we calculate the exchange and the Dzyaloshinskii-Moriya interaction as well as 
the magnetocrystalline anisotropy energy. 
The later two terms are driven by spin-orbit coupling and 
significantly reduced compared to Pd/Fe bilayers on Ir(111)
as expected since Rh and Ir are isoelectronic $4d$ and $5d$ transition-metals. 
However, there is still a spin spiral ground state at zero magnetic field. 
Atomistic spin dynamics simulations show that a skyrmion phase occurs 
at small magnetic fields of $\sim$ 1 T. Skyrmion diameters amount to 2 to 8
nm and  skyrmion lifetimes are up to 1 hour at temperatures of 25 to 45 
K.
\end{abstract}

\maketitle

Magnetic skyrmions, localized topologically protected spin structures, are promising candidates in the field of spintronics and magnetic information storage due to their unique properties~\cite{Fert2013,Wiesendanger2016,Kiselev2011}. Based on 
micromagnetic and phenomenological model calculations, skyrmions have been widely investigated in magnetic materials since the late 1980s.~\cite{Bogdanov1989,Bogdanov1994,Bogdanov:2001aa,Rossler2006} However, the experimental observation of skyrmions was achieved only very recently in \ce{MnSi} bulk material.~\cite{Muhlbauer915} Since then, researchers have shown that skyrmions can be stabilized experimentally in various systems such as bulk crystals,~\cite{Wilhem2011,Munzer2010} crystal thin films,~\cite{Yu2010b,Tonomura2012} magnetic multilayers,~\cite{Soumyanarayanan2017,Moreau-Luchaire2016} or ultrathin films.~\cite{Heinze2011,Romming2013,Romming2015}

To date, an atomic Pd/Fe bilayer on the Ir(111) surface
is the best studied ultrathin film system with a spin spiral ground state and formation of a nanoscale skyrmion lattice in an external magnetic 
field.~\cite{Romming2013,Dupe2014,Simon:14.1,Romming2015,Hanneken:15.1,Hagemeister:15.1,Leonov2016,Kubetzka:17.1} 
Skyrmions with small diameters of about 4--6 nm at magnetic fields of 1--3 T have been observed using low temperature spin-polarized scanning tunneling microscopy (STM).~\cite{Romming2013,Romming2015}
The key element of stable N\'eel-type skyrmions with a unique rotational sense in this ultrathin film is the Dzyaloshinskii-Moriya interaction (DMI).~\cite{Dzyaloshinskii1957,Moriya1960}
The DMI occurs due to strong spin-orbit coupling (SOC) from the 5$d$ transition metal (TM) with the broken structural inversion symmetry
at the interface. 
A critical value of DMI ($D_c$) is needed to stabilize noncollinear spin structures.~\cite{Bogdanov1994,Rossler2006} 
The value of $D_c$ depends on spin stiffness ($A$) and effective magnetocrystalline anisotropy constant ($K$) and is given by $D_c \propto \sqrt{AK}$. 
To stabilize spin spirals or skyrmion lattices, the DMI has to be larger than $D_c$. One approach to achieve this goal is to create interfaces of 
$3d$ and 5$d$ TM layers with large SOC. Another approach is to consider 3$d$/4$d$ interfaces to reduce the value of $K$ and thus the value of $D_c$. 
However, since DMI is also reduced for a 3$d$/4$d$ interface due to a lower SOC, very good control of magnetic interactions is needed for this approach to be successful.
Recently Herv\'{e} {\etal}~\cite{Herve2018} have reported experimental realization of skyrmions with radii of about 40--50 nm in Co monolayers on Ru(0001) in the limit of almost vanishing magnetic anisotropy.  

Here, we study an atomic Pd/Fe bilayer on the Rh(111) surface using density functional theory (DFT), atomistic spin dynamics and harmonic transition state theory. 
This system is similar to Pd/Fe/Ir(111) since Rh is the 4$d$ TM isoelectronic to Ir. 
Moreover, Fe/Rh(111) pseudomorphic films have already been experimentally grown.~\cite{Kronelein2018} 
We find that the exchange interactions are very similar to Pd/Fe/Ir(111) 
while the DMI and
$K$ are significantly reduced since SOC is weaker in Rh. However, the magnetocrystalline anisotropy energy (MAE) is affected much more strongly than the DMI 
because it is a second-order perturbation term in SOC while the DMI arises in first order. This leads to a spin spiral 
ground state at zero magnetic field as in Pd/Fe/Ir(111) and with a very similar period. 
From DFT we parameterize an extended Heisenberg model which we study by atomistic spin dynamics.
The zero temperature phase diagram shows transitions between the spin spiral ground state, skyrmion lattice, and ferromagnetic 
state under external magnetic fields on the order of 1~T. The skyrmion diameters of isolated skyrmions amount to 2 to 8~nm.
The calculated energy barriers and lifetimes for collapse of isolated skyrmions demonstrate their stability at up to 25~K to 45~K depending on Pd overlayer stacking.
%


\begin{figure}
	\centering
	\includegraphics[scale=0.7,clip]{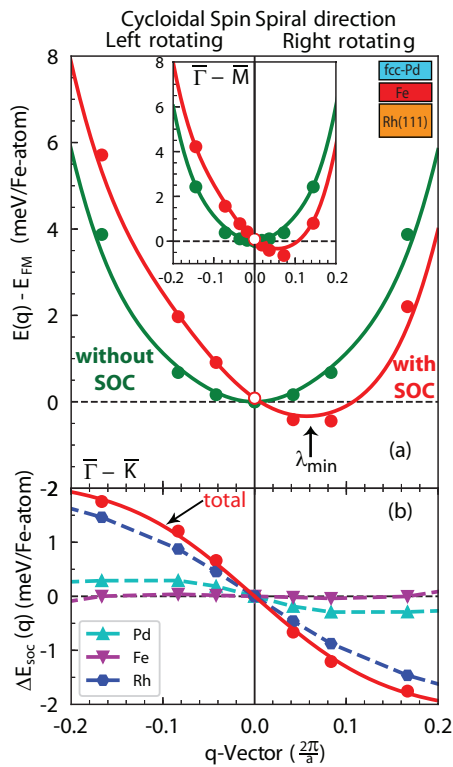}
	\caption{(a) Energy dispersion E(\textbf{q}) of homogeneous cycloidal flat spin spirals for fcc-Pd/Fe/Rh(111) without (green dots) and with SOC (red dots) in $\overline{\Gamma}-\overline{\mathrm{K}}$ high-symmetry direction for both senses of rotation. The energy is given relative to the ferromagnetic state. The dispersion is fitted to the Heisenberg model (green line) and includes the DMI and MAE (red line). Inset is the energy dispersion along the $\overline{\Gamma}-\overline{\mathrm{M}}$ direction. (b) Total and element resolved contributions of $\Delta E_{\rm SOC}(\mathbf{q})$ upon including SOC.}
	\label{fig:dispersion_fcc}
\end{figure}

We have used DFT as implemented in the \textsc{fleur} code~\cite{FLEUR} to study the electronic and magnetic properties of Pd/Fe/Rh(111). 
We have used the fcc stacking of the Fe ML on Rh(111). 
For the Pd overlayer, we have considered both fcc and hcp stacking. In order to investigate the ground state and magnetic interactions in this system, we calculated the total energy $\mathrm{E}(\mathbf{q})$ of homogeneous, flat spin spirals along the high symmetry directions $\overline{\Gamma}$--$\overline{\mathrm{K}}$ and
$\overline{\Gamma}$--$\overline{\mathrm{M}}$ in the two-dimensional Brillouin zone (2D BZ) with and without SOC.~\cite{Kurz2004,Heide2009,Zimmermann2014} Spin spirals are characterized by the wave vector $\mathbf{q}$ in the 2D BZ with a constant angle ($\phi=\mathbf{q}\cdot\mathbf{R}$) between two magnetic moments 
at adjacent lattice sites separated by $\mathbf{R}$. We calculated the 
MAE using the force theorem starting from a
self-consistent scalar-relativistic calculation and constraining the spin quantization axis along in-plane and out-of-plane directions in calculations
with SOC (see Supplemental Material for details).

Fig.~\ref{fig:dispersion_fcc}(a) displays the calculated energy dispersion $\mathrm{E}(\mathbf{q})$ of spin spirals along the $\overline{\Gamma}-\overline{\mathrm{K}}$ high-symmetry direction for fcc-Pd/Fe/Rh(111).
First, we discuss the results neglecting SOC. We observe that the energy dispersion has a minimum at the $\overline{\Gamma}$ point, i.e.~the
ferromagnetic (FM) state, and is degenerate for right-($q > 0$) and left-rotating ($q<0$) spirals. For larger values of $|q|$, the energy rises due to the 
FM nearest-neighbor (NN) exchange interactions. A similar energy dispersion is obtained for the $\overline{\Gamma}-\overline{\mathrm{M}}$ direction 
(inset of Fig.~\ref{fig:dispersion_fcc}(a)) and for hcp-Pd/Fe/Rh(111) (see Fig.~S1).  

For both Pd stackings we obtain an easy magnetization axis along the out-of-plane direction for the ferromagnetic state due to SOC. 
The values of the MAE are $K=-0.17$ and $-0.31$~meV/Fe-atom for fcc and hcp stacking, respectively.
The inclusion of SOC for spin spirals induces DMI in this system due the broken inversion symmetry at the 3$d$/4$d$ interface.~\cite{Dzyaloshinskii1957,Moriya1960} 
DMI leads to 
noncollinear spin structures with canting of the magnetic moments. The energy contributions due to the DMI along with the element decomposition are shown in Fig.~\ref{fig:dispersion_fcc}(b) for fcc-Pd/Fe/Rh(111). We observe that the major DMI contributions stem from the Rh surface together with a minor contribution from the Pd overlayer. Incorporating the DMI, the energy dispersion has a minimum value for a homogeneous cycloidal flat spin spiral state with a specific rotational sense.~\cite{Bode2007} 
As can be observed in Fig.~\ref{fig:dispersion_fcc}(a), an energy minimum of $-0.33$~meV/Fe-atom compared to the FM state occurs for a right rotating spin spiral with a pitch of $\lambda = 2\pi/q$ = 4.8~nm. 
Note that there is an energy shift of $K/2$ for spin spirals with respect to the ferromagnetic state due to the MAE 
which favors a collinear magnetic state. 
For hcp-Pd/Fe/Rh(111) we find an energy minimum of $-0.22$~meV/Fe-atom for a right rotating spin spiral with a pitch $\lambda$ = 6.7 nm. (see Fig.~S1).

We use the atomistic spin model given by
\begin{align}
H 
&= - \sum_{ij}J_{ij}(\mathbf{m}_i\cdot\mathbf{m}_j) - \sum_{ij}\mathbf{D}_{ij}\cdot(\mathbf{m}_i \times \mathbf{m}_j) \nonumber \\
&\quad+ \sum_i K(\mathrm{m}^z_i)^2 - \sum_i\mathrm{\mu_s}(\mathbf{B}\cdot\mathbf{m}_i). 
\label{eq:hamiltonian}
\end{align}
to analyze the magnetic properties of Pd/Fe/Rh(111). 
Eq.~(\ref{eq:hamiltonian}) describes the
magnetic interactions of spins 
$\mathbf{M}_i$ and $\mathbf{M}_j$ between two Fe atoms at sites $\mathbf{R}_i$ and $\mathbf{R}_j$, respectively, 
where $\mathbf{m}_i = \mathbf{M}_i/\mu_s$. 
The parameters for the exchange interactions ($J_{ij}$) and the DMI ($\mathbf{D}_{ij}$) are extracted for both fcc and hcp stacking of the Pd overlayer from the fitting of DFT energy dispersion curves calculated without SOC and with SOC, respectively. The magnetic moments ($\mathrm{\mu_s}$) and MAE ($K$) are also calculated using DFT for both stackings. The parameters $J_{ij}$, $\mathbf{D}_{ij}$, $K$, and $\mu_s$ are given in Table~\ref{table:parameters}. 
\begin{table*}
	\centering
	\caption{Exchange constants ($J_{ij}$), Dzyaloshinskii-Moriya interaction parameters ($\mathbf{D}_{ij}$), magnetocrystalline anisotropy ($K$) and total magnetic moment ($\mu_s$) obtained from DFT for 
  fcc- and hcp-Pd/Fe/Rh(111). The positive sign of $\mathbf{D}$ denotes right rotating spin spirals. The negative sign of $K$ indicates an easy out-of-plane magnetization direction. The parameters are given in meV.}
	\begin{ruledtabular}
		\begin{tabular}{c c c c c c c c c c c c | c c c c c | c | c}
			& $J_1$ & $J_2$ & $J_3$ & $J_4$ & $J_5$ & $J_6$ & $J_7$ & $J_8$ & $J_9$ & $J_{10}$ & $J_{11}$ & $D_1$ & $D_2$ & $D_3$ & $D_4$ & $D_5$ & $K$ & $\mu_s$ \\ 
			\hline
			fcc & $+13.35$ & $-2.69$ & $-2.84$ & $+0.62$ & $+0.46$ & $-0.10$ & $-0.30$ & $-0.11$ & $+0.03$ & $+0.17$ & $-0.06$ & $+0.62$ & $-0.04$ & $+0.04$ & $+0.00$ & $+0.05$ & $-0.17$ & $3.2$ \\ 
			hcp & $+12.23$ & $-1.18$ & $-2.78$ & $+0.27$ & $+0.40$ & $+0.03$ & $-0.14$ & $-0.10$ & $+0.00$ & $+0.12$ & $-0.06$ & $+0.87$ & $+0.02$ & $-0.09$ & $+0.02$ & $+0.04$ & $-0.31$ & $3.1$ \\ 
		\end{tabular}
	\end{ruledtabular}
	\label{table:parameters}
\end{table*}

The exchange constants suggest a strong frustration due to competition of ferromagnetic NN exchange and antiferromagnetic for second and
third NN exchange interactions. 
The values and their variation with the Pd overlayer stacking are quite similar to those reported for Pd/Fe/Ir(111)~\cite{VonMalottki2017a}. 
This is due to Rh being the isoelectronic to Ir which leads to a similar hybridization with the Fe layer. 
Since the spin-orbit coupling constant $\xi$ strongly increases with nuclear charge the DMI and the MAE are much smaller for Pd/Fe bilayers
on Rh(111) than on Ir(111). However, the MAE is reduced much more since it arises in second order perturbation theory with $\xi$ while the DMI occurs in first order.
The ratios of the MAE and the DMI between the two systems are quite consistent with this expectation. Taking the values of table~\ref{table:parameters}
and Ref.~\onlinecite{VonMalottki2017a} we find for the MAE $K_{\rm Pd/Fe/Rh}/K_{\rm Pd/Fe/Ir} \approx$ $0.39$ and $0.24$ and for
the DMI $D_{\rm eff}^{\rm Pd/Fe/Rh}/D_{\rm eff}^{\rm Pd/Fe/Ir}\approx 0.73$ and $0.62$ for hcp and fcc stacking, respectively.
Deviations from the simple scaling can be due to the contribution of both the Pd and the Rh (Ir) interface to the total DMI.

\begin{figure}
	\centering
	\includegraphics[scale=0.8,clip]{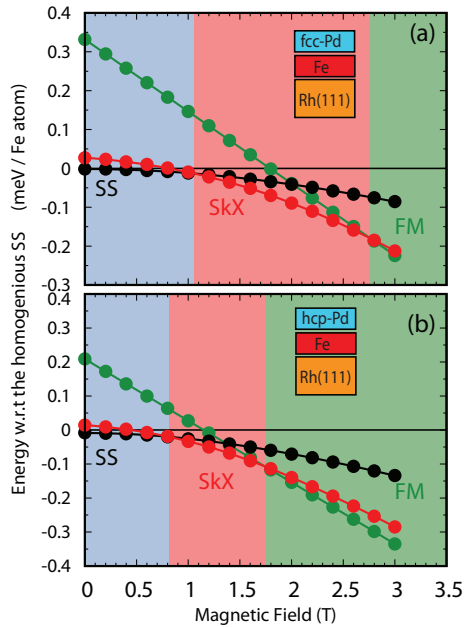}
	\caption{Zero temperature phase diagram for (a) fcc and (b) hcp Pd overlayer stacking on Fe/Rh(111) obtained based on DFT parameters for the magnetic
  interactions. The energies of the FM, skyrmion lattice (SkX) and spin spiral (SS) states are shown relative to the homogeneous spin spiral (zero line). Blue, red, and green color represents the regime of the SS, SkX, and FM ground state, respectively.}
	\label{fig:stability_fcc_hcp}
\end{figure}
\begin{figure}
    \centering
    \includegraphics[scale=0.75,clip]{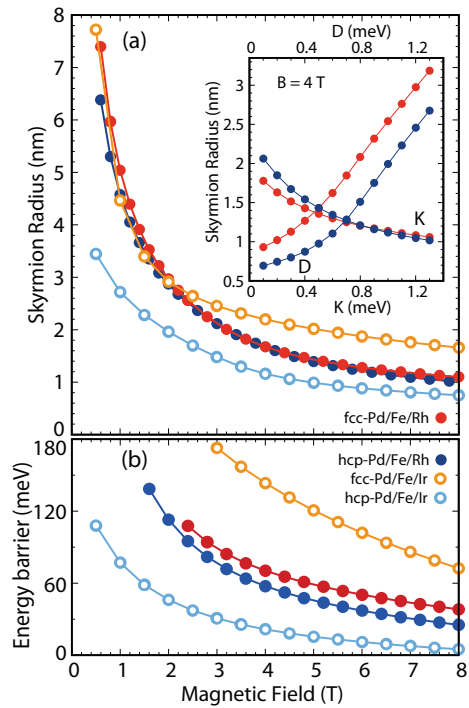}
    \caption{(a) Radii of skyrmions in Pd/Fe/Rh(111) obtained for DFT parameters as a function of the applied magnetic field. The radii of the skyrmions were computed as in Ref.~\onlinecite{Bogdanov1994b}. Inset shows the radii variations as a function of K and D for a fixed B = 4.0 T. (b) The energy barriers of isolated skyrmion collapse based on DFT parameters are shown as a function of applied magnetic field for Pd/Fe/Rh(111). For comparison the radii and energy barriers  for Pd/Fe/Ir(111) are also shown~\cite{VonMalottki2017a}.
}
\label{fig:radius_dft}
\end{figure}
To investigate the zero temperature magnetic phase transitions in Pd/Fe/Rh(111) in the presence of an external magnetic field, we have used atomistic spin-dynamics simulations using the
model 
described by Eq.~(\ref{eq:hamiltonian}) (see Supplemental Material). Fig.~\ref{fig:stability_fcc_hcp}(a,b) display the zero temperature phase diagrams   
using the parameters extracted from DFT for fcc- and hcp-Pd/Fe/Rh(111), respectively. 
For both stackings, the ground 
state at zero applied magnetic field is a 
spin spiral consistent with the energy minimum and
the pitch $\lambda$ observed in the DFT energy dispersion curves (cf.~Fig.~\ref{fig:dispersion_fcc} and Fig.~S1). 
At a certain critical field value of $\approx$ 1.06 T (0.82 T) for fcc (hcp) stacking, the skyrmion lattice is energetically favorable. 
For a larger critical field value of $\approx$ 2.76 T (1.74 T) for fcc (hcp), the skyrmion lattice phase changes to the FM phase.
These values are on the order of the fields at which skyrmions have been observed experimentally for Pd/Fe/Ir(111) \cite{Romming2013,Romming2015}.


In our simulation, isolated skyrmions are metastable in the FM phase.
We obtain their profiles 
starting from a theoretical profile~\cite{Bogdanov1994b} and then relaxing the spin structure using spin dynamics. 
For both fcc and hcp stacking, we observe skyrmions with radii 
$\approx$ 3--5 nm for magnetic fields of 1--2 T  (see Fig.~\ref{fig:radius_dft}(a)). The radii decrease rapidly with increasing value of applied magnetic field
and are larger for fcc than for hcp stacking. 
These values are very similar to those observed in Pd/Fe/Ir(111) from experiment~\cite{Romming2013}
and theory.~\cite{VonMalottki2017a} 
This may seem surprising at first glance since SOC is much smaller for Rh and both DMI and MAE are reduced significantly in our system.
In order to understand the origin of similar radii values, we have computed the radii for isolated skyrmions with given DFT exchange constants and 
either with the fixed MAE value but varying the nearest-neighbor DMI value or with fixed DMI from DFT but varying the MAE.
As shown in the inset of Fig.~\ref{fig:radius_dft}(a), the skyrmion radius decreases with reduced DMI but it rises with reduced MAE.
This observation is consistent with micromagnetic calculations.~\cite{Wang2018} 
Since both, DMI and MAE, are reduced for Pd/Fe on Rh(111) vs.~on Ir(111) the skyrmion radii remains almost unchanged.

\begin{figure}
  \centering
    \includegraphics[scale=0.7]{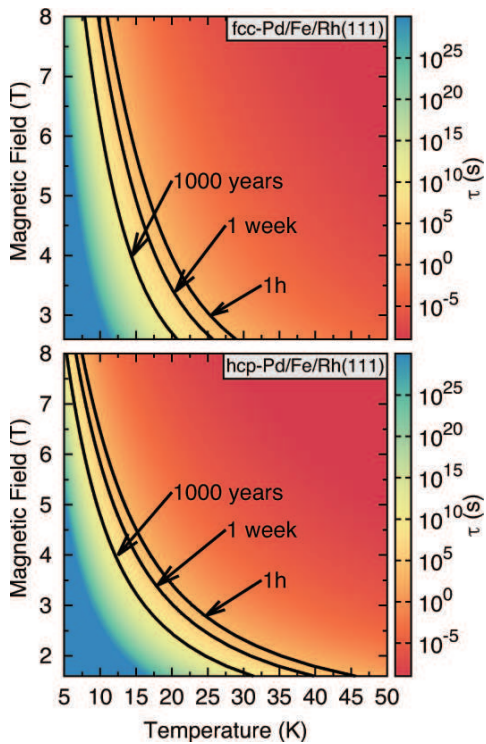}
	\caption{Lifetime of skyrmion collapse in (top) fcc-Pd/Fe/Rh(111) and (bottom) hcp-Pd/Fe/Rh(111) obtained in harmonic transition state theory
  based on DFT parameters as a function of magnetic field and temperature.
  }
	\label{fig:lifetime_fcc}
\end{figure}  

We have calculated the minimum energy path for the collapse of isolated skyrmions in the FM background 
using the geodesic nudged elastic band (GNEB) method.~\cite{Bessarab2015}
The obtained energy barriers 
[Fig.~\ref{fig:radius_dft}(b)] decrease with magnetic field and display a nonlinear 
behavior for both stackings of Pd on Fe/Rh(111).
The values of the energy barriers are in between those reported for fcc and hcp Pd on Fe/Ir(111).~\cite{VonMalottki2017a} 
Note that using the effective nearest-neighbor model (see Supplemental Material)
leads to a large reduction of the obtained energy barriers and reduced skyrmion stability as previously reported.~\cite{VonMalottki2017a} 

The decomposition of the energy barrier for skyrmion collapse in both fcc and hcp Pd stacking with respect to different interactions [see Fig.~S4]
shows that the DMI is mainly responsible for skyrmion stabilization. However, we also find
that the exchange energy leads to an additional energy barrier which is caused by exchange frustration (cf.~table~\ref{table:parameters})
and is larger than for hcp Pd stacking (see Fig.~S4). 
This automatically leads to higher energy barrier heights observed for fcc than for hcp stacking in Fig.~\ref{fig:radius_dft}(b) at a given magnetic field. 
The effective exchange parameters 
fail to describe the exchange energy barrier properly (see Fig.~S5).
The mechanism of skyrmion collapse is similar to that reported earlier.~\cite{Bessarab2015,Rohart2016,Bessarab2017prb,Rohart2017prb,VonMalottki2017a,Lobanov2016,Stosic2017,Uzdin2017,Uzdin2018} 
The skyrmion starts to shrink along the path and at the saddle point all the spins are 
nearly planar. The topological charge vanishes at this point and the skyrmion 
gradually collapse into the FM state. 

The skyrmion lifetime $\tau$ depends exponentially on the energy barrier, $\Delta E$, and temperature, $T$, 
as $\tau = \tau_0 \; 
\exp{(\Delta E / k_B T)}$.~\cite{Bessarab2012} Typically, the pre-factor $\tau_0$ is assumed to be a constant estimated value, which in some cases could be too crude approximation.~\cite{Krause2009,Bessarab2013,Wild2017,Bessarab2018}
We have explicitly calculated $\tau_0$ within harmonic transition state theory \cite{Bessarab2012,Bessarab2018} and
thereby $\tau$ as a function of temperature and external magnetic field 
[Fig.~\ref{fig:lifetime_fcc}].  
We find that isolated skyrmions should be stable in fcc-Pd/Fe/Rh(111) and hcp-Pd/Fe/Rh(111) up to hours at temperatures of up to 25~K and 45~K, respectively. Hence, these skyrmions can be observed experimentally {\eg} in STM experiments. 
However, STM
experiments on skyrmions in ultrathin films have only been performed at low temperatures of about 8~K \cite{Romming2013,Romming2015,Hagemeister:15.1,Kubetzka:17.1}
and the stability of skyrmions in these systems remains an open experimental issue. We predict a finite skyrmion lifetime 
for both fcc and hcp [cf.~Fig.~\ref{fig:lifetime_fcc}] stacking of the Pd overlayer on Fe/Rh(111) making experiments on these systems promising.

In conclusion, we predict the formation of stable isolated skyrmions with diameters of a few nanometers at a $3d/4d$ transition-metal interface. 
Our approach starts from density functional theory to obtain first-principles parameters of an atomistic spin model which is solved by
spin dynamics simulations. Using the geodesic nudged elastic band method and transition state theory enables us to obtain quantitative
results for the stability of isolated skyrmions, i.e.~energy barriers as well as lifetimes. This allows us to explore novel skyrmion systems 
which may guide future experimental work.

This work was supported by the European Unions Horizon 2020 research and innovation program under grant agreement No 665095 (FET-Open project MAGicSky). S.J.H.
and S.H. acknowledge the DFG via SFB677 for financial support. P.F.B. acknowledges support from the Icelandic Research Fund (Grants No. 163048-053, 184949-051). We gratefully acknowledge the computing time at the supercomputer of the North-German Supercomputing 
Alliance (HLRN).

%

\end{document}